\documentclass[aps,pre,preprint,nofootinbib,superscriptaddress,showpacs,showkeys]{revtex4}
\usepackage{epsfig}

\newcommand{\comment}[1]{}

\begin{document}
\preprint{APS preprint}
\title{Self-Similar Log-Periodic Structures in Western Stock Markets from 2000}
\thanks{Colour Manuscript:
${\tt http://www.physics.adelaide.edu.au/cssm/econophysics/lp\_structures\_col.ps}$}

\author{M. Bartolozzi}
\email{mbartolo@physics.adelaide.edu.au}
\affiliation{Special Research Centre for the Subatomic 
Structure of Matter (CSSM), University of Adelaide, Adelaide, SA 5005,
Australia}
\author{S. Dro\.zd\.z}
\affiliation{Institute of Nuclear Physics, Radzikowskiego 152,
31--342 Krak\'ow, Poland}
\affiliation{Institute of Physics, University of Rzesz\'ow,
PL-35-310 Rzesz\'ow, Poland} 
\author{D. B. Leinweber}
\affiliation{Special Research Centre for the Subatomic 
Structure of Matter (CSSM), University of Adelaide, Adelaide, SA 5005,
Australia}
\author{J. Speth} 
\affiliation{Institute f\"ur Kernphysik, Forschungszentrum J\"ulich,
D--52425 J\"ulich, Germany} 
\author{A. W. Thomas}
\affiliation{Jefferson Laboratory, 12000 Jefferson Ave., Newport News,
  VA 23606, USA}
\affiliation{Special Research Centre for the Subatomic 
Structure of Matter (CSSM), University of Adelaide, Adelaide, SA 5005,
Australia}

\date{\today}

\begin{abstract}
The presence of log-periodic structures before and after stock market 
crashes is  
considered to be an imprint of an intrinsic discrete scale invariance (DSI) 
in this complex system. The fractal framework of the theory  leaves open the
possibility of observing self-similar log-periodic  structures at
different time scales.  In the present work we analyze the daily closures
of four of the most important indices worldwide since 2000: the DAX for
Germany and the Nasdaq100, the S\&P500 and the Dow Jones for the United States. 
The qualitative behaviour of these different markets is similar
during the temporal frame studied.  Evidence is found for decelerating
log-periodic oscillations of duration about two years
and starting in September 2000.  Moreover, a nested
sub-structure starting in May 2002 is revealed, bringing  more
evidence to support the hypothesis of self-similar, 
log-periodic behavior.  Ongoing
log-periodic oscillations are also revealed.  A Lomb analysis over the
aforementioned periods indicates a preferential scaling factor $\lambda \sim
2$. Higher order harmonics are also present. The spectral pattern of the 
data has been found to be similar to that of a Weierstrass-type function,
used as a prototype of a log-periodic fractal function.
\end{abstract}

\keywords{Discrete Scale Invariance, Econophysics, Complex Systems}
\pacs{05.45.Pq, 52.35.Mw, 47.20.Ky}
\maketitle
\section{Introduction}
\label{}

It is well known that  many physical systems undergo phase
transitions around specific critical points in the parameter
space~\cite{stanley,sornette_cf}.  Near these points the system is
strongly correlated and many characteristic quantities  can well be
approximated by power laws, related to the scale-invariance of the
system in that state.  If we assume that $\phi(t)$ is an observable
near a critical point, $t_c$, as for example the susceptibility of the
Ising model near the critical temperature, for a change of scale $t
\rightarrow \lambda t$ we have
\begin{equation}
\phi(\lambda t) = \mu \phi(t),
\label{SI}
\end{equation}
where $\mu = \lambda^{\alpha}$ since $\phi(t) \sim
t^{\alpha}$.  The power law is a solution of Eq.(\ref{SI}) for
$\forall \lambda$.

A weaker version of the over mentioned scale invariance is the {\em
discrete-scale  invariance} (DSI) ~\cite{sornette98}.  In this case the
system becomes self-similar only for an infinite but countable set of
values of the parameter $\lambda$. That is Eq.(\ref{SI}) holds only
for $\lambda=\lambda_{1},\lambda_{2}...$ where in general
$\lambda_{n}=\lambda^{n}$.  In this case $\lambda$ represents a
preferential scaling factor that characterizes a hierarchical structure
in the system.   The solution of Eq.(\ref{SI}) can be written in a
more generic form that accounts also for a possible discrete scale
invariance:
\begin{equation}
\phi(t) = t^{\alpha} \ \Theta \left (\frac{\ln(t)}{\ln(\lambda)} \right ),
\label{DSI}
\end{equation} 
where $\Theta$ is an arbitrary periodic function of period 1.  Using a
first order Fourier expansion on Eq.(\ref{DSI}) and  writing $t
\rightarrow |t_c-t|$  we obtain
\begin{equation}
\phi(t) =A+B
|t_c-t|^{\alpha}+C|t_c-t|^{\alpha}\cos(\omega\ln|t_c-t|-\varphi),
\label{LP}
\end{equation} 
where $\omega=2\pi/\ln(\lambda)$.  The dominant power law behaviour,
a hallmark of all critical phenomena, and the log-periodic corrections to
the leading term are the main features of Eq.~(\ref{LP}).

Sornette, Johansen and Bouchaud~\cite{sornette96,sornette97} first  
pointed out how different price indices in the stock market show a
power law increase with superimposed accelerating oscillations just
before a crash.  The remarkable fact was that the log-periodic formula
(\ref{LP}), derived for DSI systems,
provided a very good approximation for this empirical
fact.  This led them to conjecture about the existence of a 
critical point in time, $t_c$,  for which the market can undergo a
phase transition (crash). 

  According to this framework the stock market is seen as a
 self-organized system that drives itself toward a critical  point.
Just as the Ising model has a parameter governing the temperature of the system
and a critical temperature where the system undergoes a phase transition,
it is postulated that the market also has such an underlying parameter which takes
the critical value at time $t_c$.
The appearance of log-periodic oscillations has been related to a
{\em discrete} hierarchal structure of the traders.
Near the critical point, when the market is very compact and
unstable, every perturbation can spread throughout  the system: a
common decision to sell by a certain group in the hierarchy of traders can
trigger a {\em herd} effect, leading to a crash.  This
concept, in a way, is similar the {\em self-organized criticality}
proposed by Bak, Tang and Wiesenfeld~\cite{SOC}, studied recently in 
Ref.~\cite{Bartolozzi04}.
 
Since the first paper by Sornette,
Johansen and Bouchaud~\cite{sornette96}  many physicists have been
attracted by the idea of phase-transitions in a self-organized stock
market~\cite{feigenbaum96,vandervalle98,drozdz99,drozdz03}, even if
criticisms have been also
raised~\cite{lalux99,ilinski99,feigenbaum01}.  A recent review on the
subject can be found  in Ref.~\cite{sornette_cc}.

An intriguing scenario has been proposed by Dro\.zd\.z and
coworkers~\cite{drozdz99,drozdz03}.  
Inspired by theoretical consistency arguments, they found 
empirical evidence that short
time log-periodic structures can be nested within log-periodic
structures on a larger scale.  The appearance of these self-similar
periods, one inside the other, has been related to the underlying
fractal nature of the DSI, giving rise to a multi-scale
log-periodicity. Moreover, the existence of a preferential scaling
ratio $\lambda \sim 2$ has been pointed out for both the leading
pattern and the related sub-structures.  This last fact led them to
formulate a universality hypothesis for the parameter $\lambda$ and as
a consequence they fixed {\em a priori} the frequency of the
oscillations to correspond to $\lambda=2$.  In this way, 
the predictive power of Eq.~(\ref{LP}) increases
considerably~\cite{drozdz99,drozdz03}.
Further evidence of embedded sub-structures has been reported recently 
by Sornette and Zhou~\cite{sornette03,zhou04}.

\begin{figure}
\vspace{1cm} \centerline{\epsfig{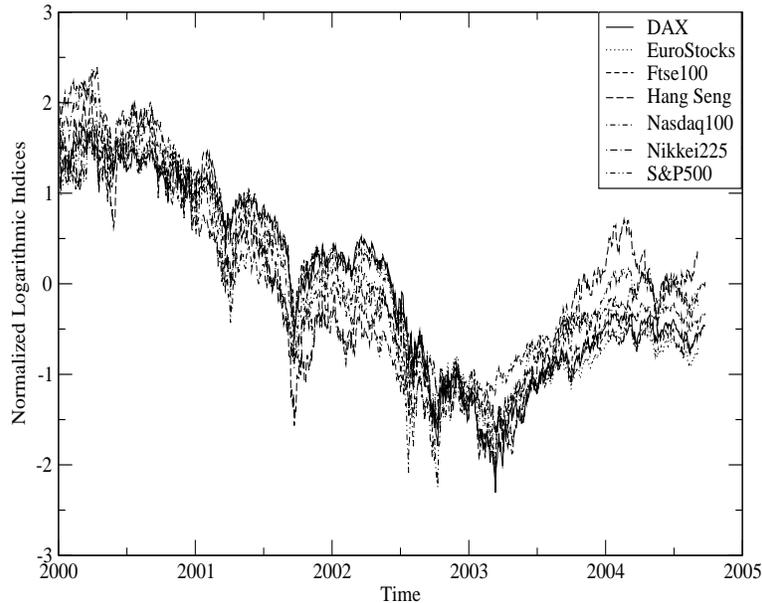}}
\caption{ Time series of the logarithm of the most important indices world-wide 
since 2000. The time series have been appropriately normalized, 
$P(t) \rightarrow \frac{P(t)-<P(t)>}{\sigma(P(t))}$, 
where $<...>$ denotes the average
over the period under consideration and $\sigma$ is the standard deviation. 
The synchronization of the indices during  the recession period,
and further on for some, is an expression of the modern globalized economy. 
The year tick in the graph marks, from now on, the position of the 1st of
January of the year itself. }
\label{fig1}
\end{figure}

Log-periodic patterns have been observed not only  in bullish periods
of the market but also during the ``antibubbles'',  or bearish periods, that
follow a market peak~\cite{johansen99,sornette02,drozdz03,zhou03}.
An example of these  log-periodic oscillations 
has been documented during the long period of recession experienced
by most of the world stock markets  since  the middle of 2000~\cite{zhou03}.
In this period, which ends approximately (in our understanding)
 in the first months of 2003,
all the most important markets world-wide are remarkably synchronized. A simple 
plot of the logarithmic indices, $P(t)$, is provided Fig.~\ref{fig1}. 
We believe this is sufficient to convince the reader of this 
behaviour. It is nothing but an expression of the growing globalization 
of the modern economy~\cite{drozdz01}.

 In the present work we focus on the study of the daily closures
 of three of the most important international indices from 
 2000 until the end of September 2004: the DAX for the
 German market and  the S\&P500, the Nasdaq100 and the Dow 
Jones\footnote[4]{The Dow Jones index has been included for historical,
more that economic, reasons. In fact, it constitutes just a small sub-set of
the S\&P500 basket but, nevertheless, it is considered by many to be 
a good indicator for the health of the economy.} 
for the American market.
 We confirm (within our DSI model of Eq.~(\ref{DSI})), the existence of a
 main log-periodic structure starting in September 2000 and ending in
 November 2002 for all the indices in Sec.~\ref{fit}.  Moreover, for
 the first time, we identify a clear bearish sub-structure, starting
 around the 15th of May 2002 and ending one year later. 
 Ongoing log-periodic oscillations with the same characteristic
 frequency, starting in January 2004, also are reported.

We also address the question of a possible universality 
of the power law exponent $\alpha$, related to the trend of the time series.
We then convert the S\&P500, Nasdaq100  and the Dow Jones in
Euros while the DAX is expressed in American Dollars.
The conversion, while leaving the oscillations unaltered, seriously distorts 
the trends.  Therefore, no universal characteristic can be
claimed for this parameter.

A Lomb analysis of the main structure and the relative sub-structure,
presented in Sec.~\ref{lomb}, reveals that the dominating frequency
of both is related to  a common value of $\lambda$ that is $\lambda
\sim 2$. A second relevant harmonic, at a frequency double that of the
fundamental, is also present.  These results  confirm the
fractal hypothesis of Dro\.zd\.z {\em at al.}~\cite{drozdz99,drozdz03}.
Further indications pointing to self-similar log-periodicity have been
found using a Lomb analysis of a  Weierstrass-type
function~\cite{berry80,gluzman02}, taken as prototype of a log-periodic
fractal function. The relevance of such a function for stock 
market log-periodic criticality was suggested for the first time
in Ref.~\cite{drozdz99}.

\section{Log-Periodicity and Sub-Structures}
\label{fit}

\begin{figure}
\vspace{1cm} \centerline{\epsfig{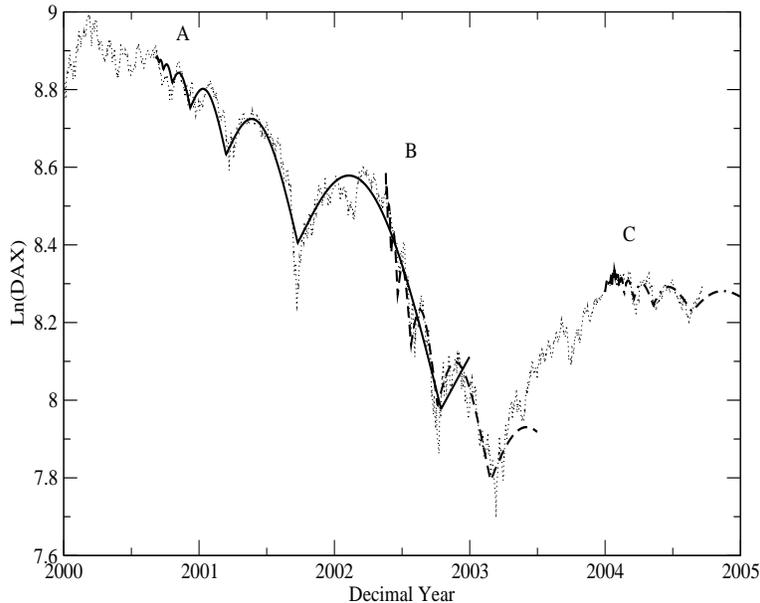}}
\caption{ Time series of the DAX from 1/1/2000 until 22/9/2004.  A
log-periodic structure of the approximate duration of two years and
starting in September 2000 is highlighted by the solid curve labeled (A).  In this
case we fix $t_c=$1/9/2000 in Eq.~(\ref{LP}). A one year
sub-structure is visible starting in May 2002, as illustrated by the dashed curve (B)
($t_c=$16/5/2002).  The dashed dotted curve labeled (C) ($t_c=$26/1/2004) is related to
the ongoing log-periodic oscillations. For all the fits we fixed
$\lambda=2$.}
\label{fig2}
\end{figure}

\begin{figure}
\vspace{1cm} \centerline{\epsfig{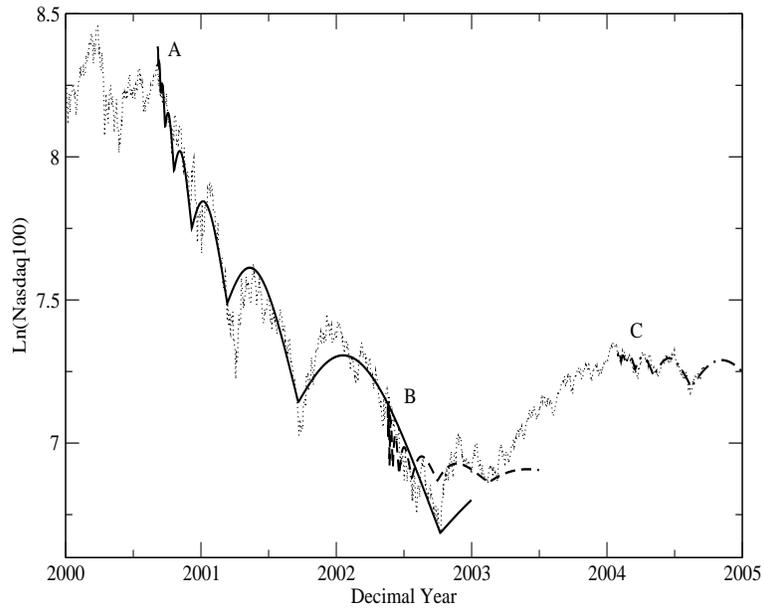}}
\caption{ Time series for the Nasdaq100 index from 1/1/2000 until
 22/9/2004.  The critical times and the coding are the same as
 in Fig.~\ref{fig2}.  The sub-structure starting in May 2002 is not as
 evident as for the S\&P500 or DAX.}
\label{fig3}
\end{figure}

\begin{figure}
\vspace{1cm} \centerline{\epsfig{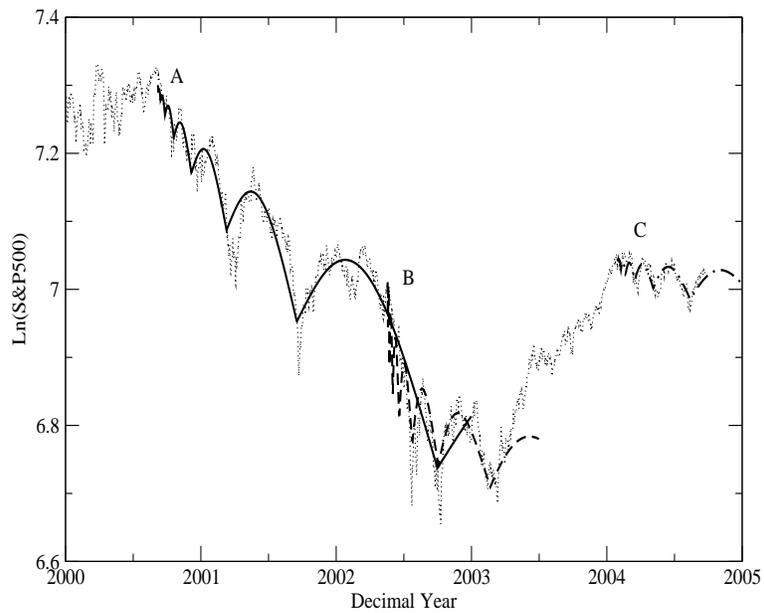}}
\caption{ Time series for the S\&P500 index from 1/1/2000 until
 22/9/2004.  The critical times and  coding are the same as in
 Fig.~\ref{fig2}.}
\label{fig4}
\end{figure}

\begin{figure}
\vspace{1cm} \centerline{\epsfig{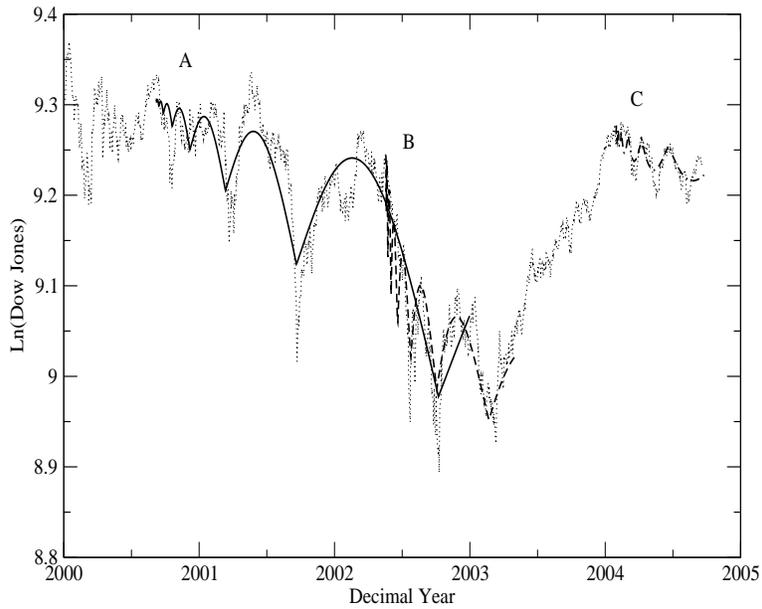}}
\caption{ Time series for the Dow Jones index from 1/1/2000 until
 22/9/2004.  The critical times and  coding are the same as in
 Fig.~\ref{fig2}.}
\label{fig5}
\end{figure}

The DSI model of Eq.~(\ref{LP}) is used to fit the logarithm of the
DAX, Nasdaq100 S\&P500 and Dow Jones.
 Considering that the periodic function $\Theta$ is
arbitrary, we take the modulus of the cosine as it provides a better
representation of the data~\cite{drozdz03} than the cosine itself.  
The fitting procedure that we use is borrowed from Ref.~\cite{sornette96}.  
According to that we express the  parameters from $A$, $B$ and
$C$ of Eq.~(\ref{LP}) as a function 
of $\alpha$, $\omega$ (or $\lambda$), $\varphi$ and
$t_c$ by imposing the constraint that the cost function, $\chi^2$,
has null derivative with respect to them. 
Following the method  by Dro\.zd\.z {\em at al.}~\cite{drozdz99,drozdz03},
we  fix the parameter $\lambda$, considering it to be a universal constant. 
In particular we choose $\lambda=2$. If this assumption turns out to be
confirmed, then the predictive power of the model will
considerably be increased. Moreover, as we are interested here  in
bearish periods only and not in predicting the most probable crash time, we
introduce a further simplification by adjusting $t_c$  visually.
In this way there are only two parameters left  to explore:
$\alpha$ and $\varphi$.
The results of the fits are shown in Figs.~\ref{fig2}, \ref{fig3}, \ref{fig4} and
\ref{fig5} for the DAX, the Nasdaq100, the S\&P500 and Dow Jones 
respectively.

A log-periodic structure starting around the 1st of September 2000 and
finishing in November 2002 is clearly evident for all the sets of
data considered. Moreover a nested sub-structure starting  around
the 15th of May 2002 is also visible.
Log-periodic oscillations also characterize  
the present state of the
market.  A possible origin of this behaviour can be localized at the
end of January  2004.

At this point it is important to emphasize that the previous fits are
obtained for a preferential scaling coefficient, $\lambda=2$. This is
already a good indication of the universal nature of the hierarchical
scaling in stock markets.

\begin{figure}
\vspace{1cm} \centerline{\epsfig{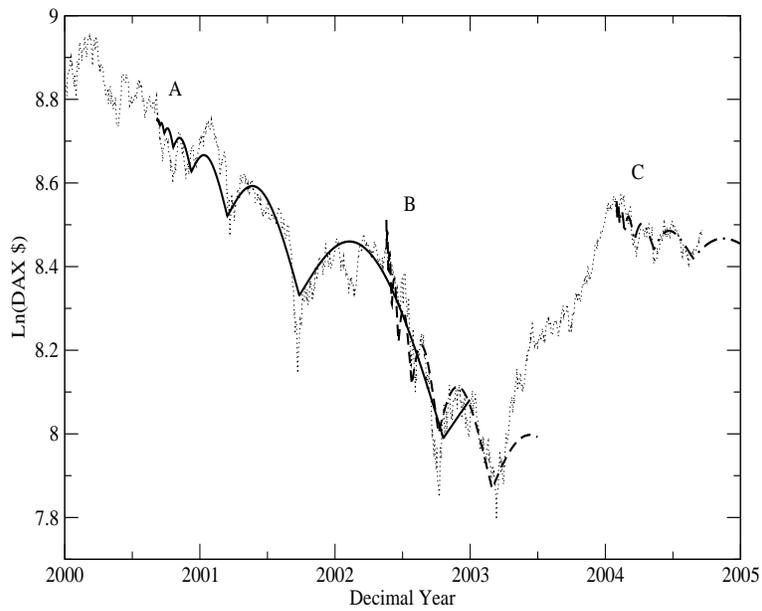}}
\caption{ The DAX expressed in
American dollars. Features are as described in Fig.~\ref{fig2}.}
\label{fig6}
\end{figure}

\begin{figure}
\vspace{1cm} \centerline{\epsfig{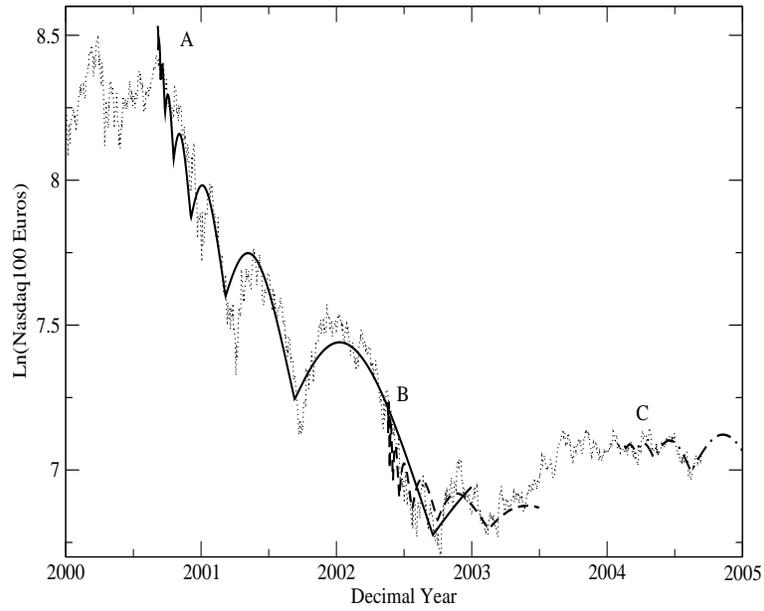}}
\caption{ The Nasdaq100 expressed
in Euros. Features are as described in Fig.~\ref{fig2}.}
\label{fig7}
\end{figure}

\begin{figure}
\vspace{1cm} \centerline{\epsfig{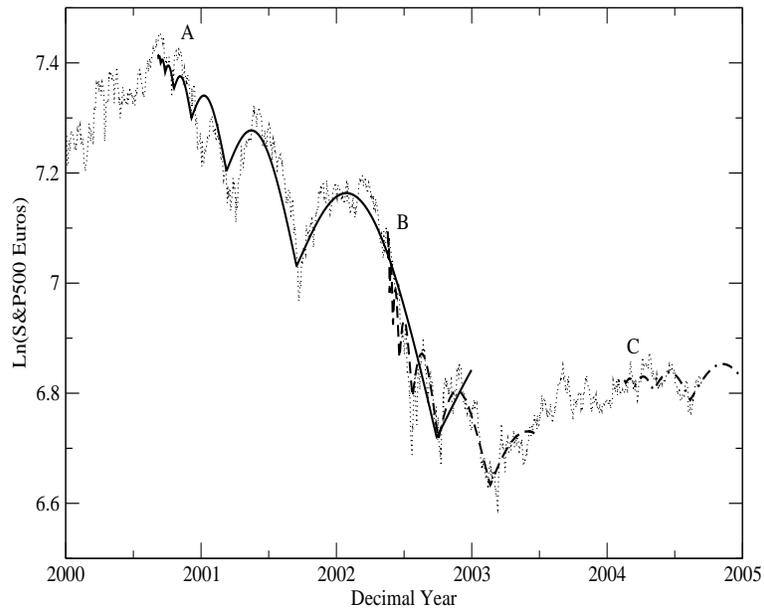}}
\caption{ The S\&P500 converted from
 American dollars to Euros. Features are as described in Fig.~\ref{fig2}.}
\label{fig8}
\end{figure}

Another interesting point regards the universality of the parameter
$\alpha$, related to the main trend of the time series.  In order to
have some insight into this direction we repeat the previous fits after
the conversion of the various indices into different currencies. That is
we transform the DAX from Euros to American dollars, and the Nasdaq100,
S\&P500 and Dow Jones from American dollars to Euros.  The results are shown in
Figs. \ref{fig5}, \ref{fig6}, \ref{fig7} and \ref{fig8}.

\begin{figure}
\vspace{1cm} \centerline{\epsfig{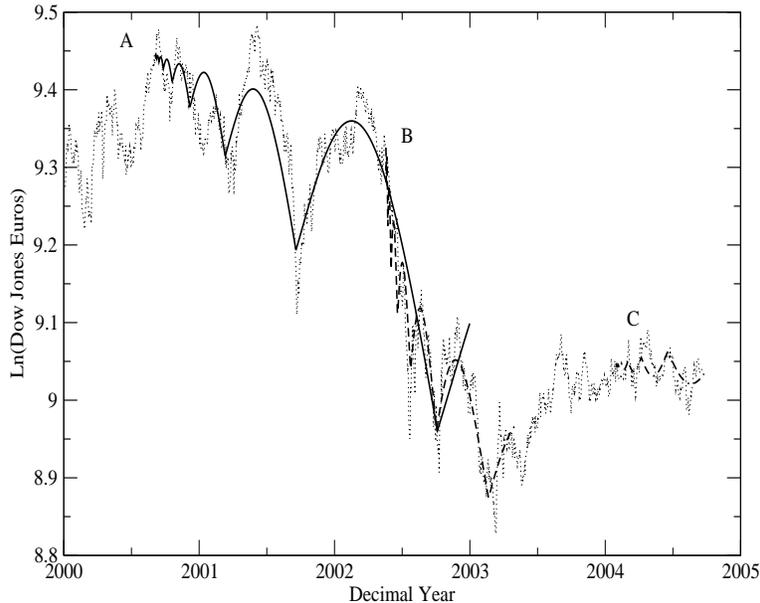}}
\caption{ The Dow Jones converted from
 American dollars to Euros. Features are as described in Fig.~\ref{fig2}.}
\label{fig9}
\end{figure}

It appears clear from the plots that, while the oscillatory structures
are unaltered by the currency conversion, the trends experience a
serious distortion and therefore we cannot extract a universal
characteristic for the  exponent $\alpha$. On the other hand, one 
might wonder to what extent the market dynamics are digested in a currency
other than the native 
currency\footnote[3]{The problem of a change in the currency of the S\&P500 
index 
has been addressed also in Ref.~\cite{zhou05}. In this case it was argued
that the main source of distortion of this index is related to the depreciation
of the American dollar due to the feedback action
of the Federal Reserve Bank.}.

\section{A Non-Parametric Approach: the Lomb Analysis}
\label{lomb}

In order to justify our assumption of $\lambda=2$ we perform a
non-parametric test on the angular frequency value of the log-periodic
oscillations. Following the method proposed by Johansen {\em et al.} in
Ref.~\cite{johansen99b}, we analyze  the time
series of residuals, $R(t)$, obtained by removing the leading power law trend
in the logarithm of the price, $P(t)$, according to
\begin{equation}
R(t) = \frac{P(t)-A-B|t_c-t|^{\alpha}}{C|t_c-t|^{\alpha}}.
\label{remove}
\end{equation} 
If the model of Eq.~(\ref{LP}) reproduces the behaviour of
the market correctly then the residual dependence on in the 
variable $\ln(|t_c-t|)$ must
be a cosine function and a spectral analysis should reveal a high peak
corresponding to the angular frequency $\omega$. 

\begin{figure}
\vspace{1cm} \centerline{\epsfig{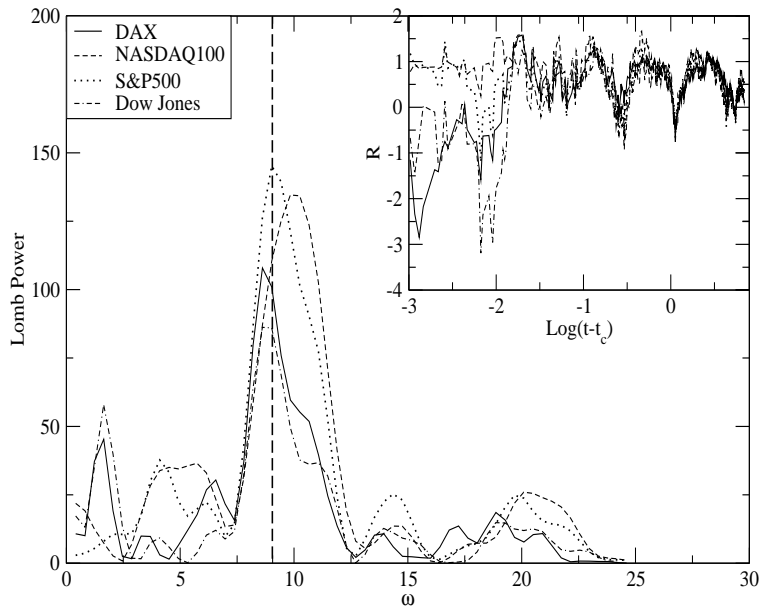}}
\caption{ Lomb analysis for the DAX (solid line), Nasdaq100 (dashed line), 
S\&P500 (dotted line)  and Dow Jones (dashed-dotted line) 
during the period from 1/9/2000 to 30/12/2002. A main
frequency around  $\tilde{\omega}=9.06$ ($\lambda=2$), dashed vertical line, is
clearly evident. Another  important harmonic contribution can be seen
at $\omega \approx 2\tilde{\omega}$.  In the insert the residuals,
with the same coding, are plotted.}
\label{fig10}
\end{figure}

Once we have obtained the residuals,  shown in the inserts of
Figs.~\ref{fig8},~\ref{fig9} and \ref{fig10}, 
we apply a spectral decomposition of
these signals according to the Lomb algorithm~\cite{press94}.  This
spectral algorithm makes use of a series of local fits using a cosine
function with a phase and provides some practical advantages, compared
to the classical Fourier transform, when the data under examination are not
evenly sampled, as in our case.
The results of the Lomb analysis  for the periods under consideration
are presented in Figs.~\ref{fig8},~\ref{fig9} and \ref{fig10}. In these
plots the angular frequency corresponding to $\lambda=2$, that is
$\omega \approx 9.06$, is represented by a vertical dashed line.
%
%
\begin{figure}
\vspace{1cm} \centerline{\epsfig{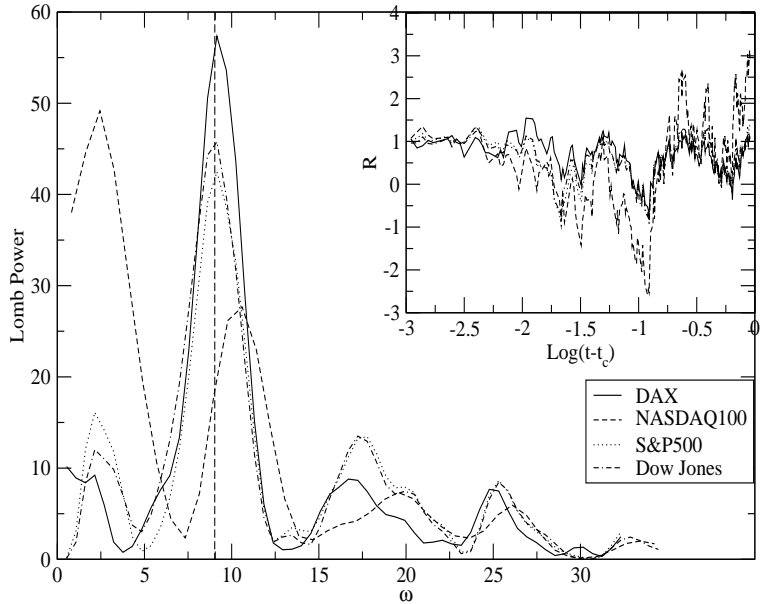}}
\caption{Lomb analysis for the DAX (solid line), Nasdaq100 (dashed line),
S\&P500 (dotted line) and  Dow Jones (dashed-dotted line)
 during the period from 16/5/2002 to 3/5/2003. Regular
high order harmonics are still present. As already seen from the fit,
the log-periodic behaviour of the Nasdaq100 index is not as clear as
for the other indices. The residuals are presented in the insert.}
\label{fig11}
\end{figure}
%
\begin{figure}
\vspace{1cm} \centerline{\epsfig{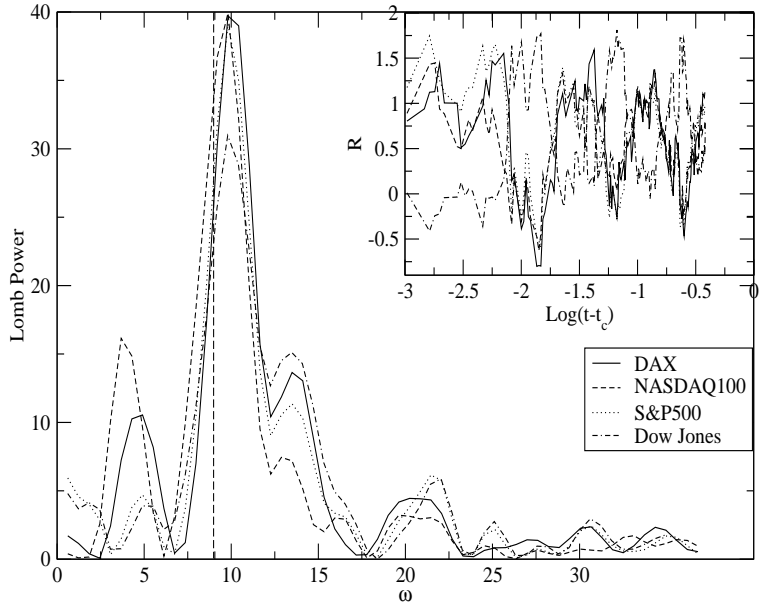}}
\caption{Lomb analysis for the DAX (solid line), Nasdaq100 (dashed line),
S\&P500 (dotted line)  and  Dow Jones (dashed-dotted line) 
during the period from 26/1/2004 to 22/9/2004. A main
frequency  along with other higher order harmonics are revealed. In this case
the phase of the residuals for the Dow Jones differs from
the other indices, reflecting a less then perfect synchronization.}
\label{fig12}
\end{figure}
%
The results of the analysis show, for all the periods, a dominating
peak  in the vicinity of $\lambda=2$.  This fact brings more evidence
to the existence of a universal  scaling factor $\lambda \approx 2$.
It is also important to underline how the Lomb analysis shows, in most
of the cases, a second main frequency that is about two times the
leading frequency.

We can also assign a confidence level to the main peaks found in the
analysis. Following the technique proposed in Ref.~\cite{zhou02}, 
the ratio between the two highest
peaks in the Lomb periodogram is used to give an estimation of the
significance level of the higher peak. 
In selecting the peaks for the ratio are excluded
 the higher-order harmonics, multiples of the fundamental.
  All the ratios found from our analysis
are clustered in a range approximately between 3 and 6, except
for the Nasdaq100 index of Fig.~\ref{fig9}. 
Even for a ratio of about 3, the confidence
 level is higher than 99\%  assuming Gaussian noise~\cite{zhou02}.
  If, instead, we assume that the noise is temporally correlated,
then a fractional Brownian motion~\cite{feder} with Hurst
exponent, $H$, at the worst, unrealistic,
case of $H=0.9$, provides a confidence level which remains greater than 
80\%~\cite{zhou02}. Hence all the peaks found by the
Lomb analysis show a high statistical significance, with the only exception
being that of the Nasdaq100 where the first peak has a low confidence level.

In order to have a better understanding of the spectral patterns
just found, we test the same method of analysis on a
Weierstrass-type function~\cite{berry80,gluzman02}.
The Weierstrass-type functions are a particular solution of the discrete
renormalization group equation for critical
phenomena~\cite{sornette_cf,gluzman02}.  Defined in the interval
[0,1], these functions are characterized by a  self-similar hierarchy
of log-periodic structures accumulating at a critical point $t_c$
($t_c$ can be 0 or 1 according to particular  choices of the
parameters).  Zhou and Sornette have shown that Weierstrass-type
functions can provide a good approximation for the bearish period of
the stock market starting from 2000~\cite{zhou03b}.

The following Weierstrass-type function~\cite{sornette_cf,gluzman02}
has been used as a test for the Lomb method:
\begin{equation}
f(t)=\sum_{n=0}^{\infty}\frac{1}{\lambda^{(2-D)n}}\exp[-\lambda^{n}t\cos(\gamma)]
\cos[\lambda^{n}t\sin(\gamma)],
\label{weierstrass}
\end{equation} 
where $\gamma \in[0,\pi/2]$ is a parameter that fixes the oscillatory
structures. If $\gamma=\pi/2$ then the parameter $D$
corresponds to the fractal dimension of the function. For
$\gamma<\pi/2$ the function becomes  smooth and it is no longer
fractal~\cite{sornette_cf,gluzman02} but preserves the large scale
log-periodic oscillation. Characteristic curves are illustrated in  Fig.~\ref{fig11}.
  
\begin{figure}
\vspace{1cm} \centerline{\epsfig{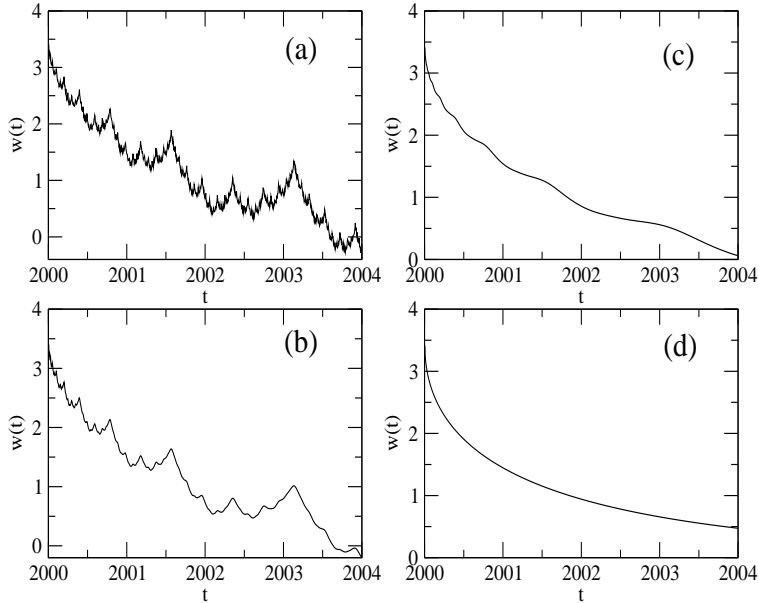}}
\caption{ The Weierstrass-type function of Eq.~(\ref{weierstrass}) for
(a) $\gamma=\pi/2$, (b) $\gamma=0.93\pi/2$, (c) $\gamma=0.90\pi/2$
and (d) $\gamma=0$. For all the plots $D=1.5$ and $\lambda=2$. The
sum in Eq.~(\ref{weierstrass})
has been truncated at $N=32$ because the function does not change
significantly beyond this value of $N$. The time axis has been
rescaled, as the Weierstrass-type function is defined only  for
$t \in [0,1]$. In the present plots the function is fractal only in (a).}
\label{fig13}
\end{figure}

Once we have chosen the test function, the same procedures  as used for
the stock market time series are applied  to the artificial
time series generated by  Eq.~(\ref{weierstrass}) with $\lambda=2$.
fixed. In this case the Lomb analysis reveals for the   fractal case
($\gamma=\pi/2$), apart from the clear peak at the main frequency, 
other smaller, high frequency harmonics regularly
spaced, as illustrated in Fig.~\ref{fig12}.

\begin{figure}
\vspace{1cm} \centerline{\epsfig{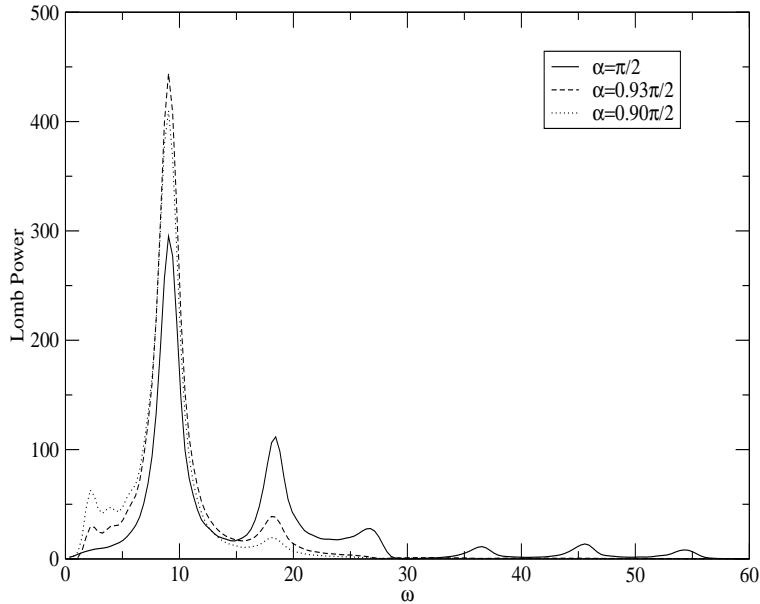}}
\caption{ Lomb analysis for the Weierstrass-type function of
Eq.(\ref{weierstrass}).  The self-similarity of the function in the
fractal case ($\gamma=\pi/2$) is reflected in the regularity of the
high order harmonics. Once we smooth the function, the high order
harmonics related to the fractality disappear while the  dominant
frequency of the log-periodic leading term is unaltered.}
\label{fig14}
\end{figure}

The higher frequency periodic peaks are a manifestation of the
fractality of the function itself. In fact, for $\gamma<1$ those
frequencies are absent.  The high frequency harmonics, in the pure
fractal case ($\gamma=\pi/2$), are  a reflection of the
self-similarity of the function at different scales.

The real data has similar high frequency modes that have
about the same spacing as the ones artificially obtained with the
Weierstrass-type function.  We can also conjecture that these modes
are related, as in the previous case, to self-similar structures in
the time series.  Of course, because of the lack of points and noise
effects, these harmonics are not as clear as for the Weierstrass-type
function.

The similarity in the spectral pattern  between real and artificial
data is an indication of the existence of self-similar structures
at different scales  in stock market time series, providing further support 
for the fractal framework of  Dro\.zd\.z and
coworkers~\cite{drozdz99,drozdz03}.  The sub-structure starting in May
2002 is a clear example of self-similarity in stock market
dynamics.

\section{Conclusions}
\label{conclusion}

In the present work we have shown that, at least, three clear
log-periodic periods have characterized, and still characterize, the
behaviour of some of the most important indices worldwide since the
year 2000.  Moreover, one of the log-periodic structures found is
embedded in  a longer one, interestingly, both in the decelerating market phase.
This finding supports the hypothesis of
self-similar log-periodicity proposed by Dro\.zd\.z and
coworkers~\cite{drozdz99,drozdz03}.  A non parametric analysis over
these periods has also been performed.  The results of the
analysis confirm the existence of log-periodic structures.
Moreover, we found further evidence for a preferential scaling factor of
$\lambda \sim 2$.  The presence of a higher order harmonic at a
frequency that is double the fundamental can also be related to
the fractal structure of the time series. A test on a Weierstrass-type
function supports this hypothesis.  We have also investigated
a possible universality of the power law index $\alpha$. For this
purpose we have converted the price time series to different
currencies, namely the DAX from Euros to American Dollars and the
Nasdaq100, the S\&P500 and the Dow Jones 
from American dollars to Euros. While   the
log-periodic oscillations remain unaltered by this procedure, the
trends come to be seriously distorted and no universality can be
claimed.

\section{Acknowledgement}
This work was supported by the Australian Research Council. J.S.
thanks Tony Thomas for the hospitality he enjoyed during his visit
to the CSSM. M. B. wishes to thank Josef Speth for his kind 
hospitality at the Forschungszentrum. We also would like
to thank D. Sornette for useful comments on the manuscript.

\end{document}